\pdfoutput=1
\documentclass[11pt]{article}

\usepackage{graphicx}
\usepackage{amsmath}
\usepackage{amssymb}
\usepackage{dcolumn}
\usepackage{bm}
\usepackage{slashed}
\usepackage{color}
\usepackage{authblk}

\newcommand{\be}{\begin{equation}}
\newcommand{\ee}{\end{equation}}
\newcommand{\ba}{\begin{eqnarray}}
\newcommand{\ea}{\end{eqnarray}}

\newcommand{\gsim}{\mathrel{\hbox{\rlap{\lower.55ex \hbox {$\sim$}}
                   \kern-.3em \raise.4ex \hbox{$>$}}}}
\newcommand{\lsim}{\mathrel{\hbox{\rlap{\lower.55ex \hbox {$\sim$}}
                   \kern-.3em \raise.4ex \hbox{$<$}}}}

\def\vk{{\vec k}}
\def\vq{{\vec q}}
\def\vp{{\vec p}}

\def\roughly#1{\mathrel{\raise.3ex\hbox{$#1$\kern-.75em%
\lower1ex\hbox{$\sim$}}}}
\def\lsim{\roughly<}
\def\gsim{\roughly>}

\def\hb{\hbar}
\def\hp{{\hat p}}
\def\hq{{\hat q}}
\def\hk{{\hat k}}
\def\({\left(}
\def\){\right)}
\def\[{\left[}
\def\]{\right]}

\def\l{{\lambda}}

\def\d{{\delta}}
\def\D{{\Delta}}

\def\O{{\Omega}}
\def\e{{\epsilon}}

\def\a{{\alpha}}
\def\b{{\beta}}
\def\c{{\chi}}

\def\g{{\gamma}}

\def\h{\eta}
\def\p{{\pi}}
\def\P{{\Pi}}
\def\m{{\mu}}
\def\n{{\nu}}
\def\r{{\rho}}
\def\s{{\sigma}}
\def\S{{\Sigma}}
\def\t{{\tau}}
\def\th{{\theta}}
\def\ph{{\phi}}
\def\ps{{\psi}}

\def\P{{\Pi}}
\def\z{{\zeta}}

\newcommand{\pd}{{\partial}}

\newcommand{\tr}{\text{tr}}

\newcommand{\tf}{\tilde{f}}

\newcommand{\tS}{\underline{S}}
\newcommand{\feq}{f^{(0)}}
\newcommand{\fneq}{f^{(1)}}
\newcommand{\tfeq}{{\tilde f}^{(0)}}
\newcommand{\tfneq}{{\tilde f}^{(1)}}

\date{\today}

\begin{document}

\title{\bf Shear induced polarization: Collisional contributions}

\author[1]{{Shu Lin}
\thanks{linshu8@mail.sysu.edu.cn}}
\affil[1]{School of Physics and Astronomy, Sun Yat-Sen University, Zhuhai 519082, China}
\author[2]{{Ziyue Wang}
\thanks{zy-wa14@mails.tsinghua.edu.cn}}
\affil[2]{Department of Physics, Tsinghua University, Beijing 100084, China}

\maketitle

\begin{abstract}

It has been realized that thermal shear plays a similar role as thermal vorticity in polarizing spin of particles in heavy ion collisions. We point out that shear has a fundamental difference that it leads to particle redistribution in the medium. The redistribution gives rise to an additional contribution to spin polarization through the self-energy, which is parametrically the same order as the one considered so far in the literature. The self-energy contribution is in general gauge dependent. We introduce double gauge links stretching along the Schwinger-Keldysh contour to restore gauge invariance. We also generalize the straight path to adapt to the Schwinger-Keldysh contour. We find another contribution associated with the gauge link, which is also parametrically the same order. We illustrate the two contributions with a massive probe fermion in massless QED plasma with shear. A modest suppression of spin polarization is found from the combined contributions when the probe fermion has momentum much greater than the temperature.

\end{abstract}


\newpage

\section{Introduction}

It has been suggested that orbital angular momentum carried by participants in off-central heavy ion collisions (HIC) can result in spin polarization of final state particles \cite{Liang:2004ph,Liang:2004xn}. Realistic model calculations have indicated that significant vorticity is present in quark-gluon plasma (QGP) produced in HIC \cite{Deng:2012pc,Pang:2016igs,Xia:2018tes}. Theoretical predictions of final particle spin polarization have been made based on a spin-orbit coupling picture \cite{Gao:2007bc,Huang:2011ru,Jiang:2016woz}. Such a picture is indeed consistent with early experimental measurement of Lambda hyperon global polarization \cite{STAR:2017ckg}. However, recent measurement of Lambda hyperon local polarization \cite{STAR:2019erd} shows an overall sign difference from theoretical predictions \cite{Becattini:2017gcx,Wei:2018zfb,Fu:2020oxj}. Different explanations have been proposed to understand the puzzle \cite{Wu:2019eyi,Liu:2019krs}, yet no consensus has been reached.

Recently it has been realized that shear can also contribute to spin polarization \cite{Liu:2021uhn,Becattini:2021suc}. In particular, it has been found based on a free theory analysis that spin responds to thermal vorticity and thermal shear in the same way. Phenomenological implementations have shown the right trend toward the measured local polarization results \cite{Fu:2021pok,Becattini:2021iol,Yi:2021ryh,Fu:2022myl,Wu:2022mkr}. However, as we shall show in this paper, the contribution discussed so far is still incomplete. Vorticity and shear differ in one important aspect: the former does not change the particle distribution while the latter necessarily does. The redistribution of particles by shear flow leads to an extra contribution to spin polarization. The extra contribution can be consistently described in the framework of quantum kinetic theory (QKT), see \cite{Hidaka:2022dmn} for a review and references therein. Rapid development of QKT has been made to include collisional term systematically via self-energy over the past few years\cite{Hidaka:2016yjf,Zhang:2019xya,Li:2019qkf,Carignano:2019zsh,Yang:2020hri,Wang:2020pej,Shi:2020htn,Weickgenannt:2020aaf,Hou:2020mqp,Yamamoto:2020zrs,Weickgenannt:2021cuo,Sheng:2021kfc,Wang:2021qnt,Lin:2021mvw}. The QKT is formulated using the Wigner function, whose axial component can be related to spin polarization. The axial component of Wigner function for fermion in a collisional QKT is given by \cite{Yang:2020hri,Lin:2021mvw,Hattori:2019ahi}\footnote{The definitions of Wigner function in \cite{Yang:2020hri} and \cite{Lin:2021mvw} differ by a sign. We use the latter definition.}
\begin{align}\label{calA}
{\cal A}^\m=-2\p\hb\[a^\m f_A+\frac{\e^{\m\n\r\s}P_\r u_\s{\cal D}_\n f}{2(P\cdot u+m)}\]\d(P^2-m^2),
\end{align}
with $P$ and $u$ being momentum of particle and flow velocity. $a^\m f_A$ is a dynamical contribution \cite{Hattori:2019ahi,Weickgenannt:2019dks,Gao:2019znl,Liu:2020flb,Guo:2020zpa}. ${\cal D}_\n$ is a covariant derivative acting on the distribution function $f$ defined as ${\cal D}_\n=\pd_\n-\S_\n^>-\S_\n^<\frac{1-f}{f}$. The partial derivative term is what has been considered so far, the extra contribution comes from self-energies $\S^{>/<}$. Naively one may expect the self-energy term to be suppressed by powers of coupling in a weakly coupled system described by the QKT. In fact this is not true. In a simple relaxation time approximation, the self-energy contribution can be estimated as $\frac{\d f}{\t_R}$. The appearance of $\d f$ follows from the fact that the self-energy contribution in the covariant derivative vanishes in equilibrium by detailed balance. The combination $\frac{\d f}{\t_R}$ can be further related to $\pd f_0$ by kinetic equation with $f_0$ being local equilibrium distribution. Consequently the self-energy contribution is at the same order as the derivative one, with the dependence on coupling completely canceled between $\frac{1}{\t_R}$ and $\d f$.

A second question we attempt to address is the gauge dependence of spin polarization. Since theoretical calculation is usually done in the QGP phase while experiments measure particle after freezeout. The gauge dependence is only present in the partonic level calculations. On general ground, we expect that it is a gauge invariant spin polarization that is passed through freezeout. However, \eqref{calA} is expressed in terms of self-energy, which is in general gauge dependent. It is necessary to include gauge link contribution to restore gauge invariance. Since collisions are mediated by off-shell particles, it is essential to consider quantum gauge field fluctuations in the gauge link. The quantum gauge field fluctuation also feels the flow via interaction with on-shell fermions. It turns out that there is a similar contribution associated with the gauge link, which is also at the same order as the derivative one. As a conceptual development, we generalize the definition of gauge link to the Schwinger-Keldysh contour, in which the collisional QKT is naturally derived. We also adapt the straight path widely used for background gauge field to the Schwinger-Keldysh contour to allow for consistent treatment of quantum gauge field fluctuations.

The aim of the paper is to evaluate the two contributions mentioned above. We illustrate the calculations by using a massive probe fermion in a massless QED plasma. While the method we use is applicable to arbitrary hydrodynamic flow, we consider the plasma with shear flow only for simplicity. The paper is organized as follows: in Section II, we briefly review the classical limit of QKT, which is the Boltzmann equation widely used in early studies of transport coefficients. By solving the Boltzmann equation we determines the particle redistribution in the presence of shear flow. The information of particle redistribution will be used to calculate the self-energy contribution and the gauge link contribution in Section III and IV respectively. Analytic results can be obtained at the leading logarithmic order. The results will be discussed and compared with the derivative contribution in Section V. Finally we summarize and provide outlook in Section VI.

\section{Particle redistribution in shear flow}

We consider a QED plasma with $N_f$ flavor of massless fermions in a shear flow. The shear flow relaxes on the hydrodynamic scale, which is much slower than the relaxation of plasma constituents, thus we can take a steady shear flow. The presence of shear flow leads to redistribution of fermions and photons, which gives rise to off-equilibrium contribution to energy-momentum tensor responsible for shear viscosity. The kinetic equation addressing this problem has been written down long ago \cite{Arnold:2002zm,Arnold:2000dr,Arnold:2003zc}. The kinetic equation is simply the Boltzmann equation with collision term given by elastic and inelastic scatterings. For simplicity we keep to the leading-logarithmic (LL) order, for which the inelastic scatterings are irrelevant. The resulting Boltzmann equations for fermion and photon read respectively
\begin{subequations}\label{Boltzmann}
\begin{align}
\(\pd_t+{\hat p}\cdot\nabla_x\)f_p=&-\frac{1}{2}\int_{p',k',k}(2\p)^4\d^4(P+K-P'-K')\frac{1}{16p_0k_0p_0'k_0'}\times\nonumber\\
&\bigg[\;\;|{\cal M}|_{\text{Coul},f}^2\(f_pf_k(1-f_{p'})(1-f_{k'})-f_{p'}f_{k'}(1-f_{p})(1-f_{k})\)\nonumber\\
&+|{\cal M}|_{\text{Comp},f}^2\(f_p\tf_k(1+\tf_{p'})(1-f_{k'})-\tf_{p'}f_{k'}(1-f_{p})(1+\tf_{k})\)\nonumber\\
&+|{\cal M}|_{\text{anni},f}^2\(f_pf_k(1+\tf_{p'})(1+\tf_{k'})-\tf_{p'}\tf_{k'}(1-f_{p})(1-f_{k})\)\bigg],
\end{align}
\begin{align}
\(\pd_t+{\hat p}\cdot\nabla_x\)\tf_p=&-\frac{1}{2}\int_{p',k',k}(2\p)^4\d^4(P+K-P'-K')\frac{1}{16p_0k_0p_0'k_0'}\times\nonumber\\
&\bigg[\;\;|{\cal M}|_{\text{Comp},\g}^2\(\tf_pf_k(1-f_{p'})(1+\tf_{k'})-f_{p'}\tf_{k'}(1+\tf_{p})(1-f_{k})\)\nonumber\\
&+2N_f|{\cal M}|_{\text{anni},\g}^2\(\tf_p\tf_k(1-\tf_{p'})(1-\tf_{k'})-f_{p'}f_{k'}(1+\tf_{p})(1+\tf_{k})\)\bigg].
\end{align}
\end{subequations}
We have used $f_p$ and $\tf_p$ to denote distribution functions for fermions and photon carrying momentum $p$ respectively. $|{\cal M}|^2$ is partially summed amplitude square with the subscripts ``Coul'', ``Comp'' and ``anni'' indicate Coulomb, Compton and annihilation processes respectively. The subscripts $f$ and $\g$ distinguish the fermionic and photonic amplitude squares, whose explicit expressions we shall present shortly. The overall factor $\frac{1}{2}$ on the RHS coming from spin average and $\int_p\equiv\int\frac{d^3p}{(2\p)^3}$. When there is imbalance between electron and position, there should be a separate equation for position. We restrict ourselves to neutral plasma, in which the positron distribution is identical to that of anti-fermion.

Now we can work out the redistribution of particles in the presence of thermal shear, given by solution to the Boltzmann equation. We solve \eqref{Boltzmann} by noting that $f_p$ and $\tf_p$ on the LHS are the local equilibrium distributions and deviations of equilibrium distributions appear only on the RHS. We can parametrize the local equilibrium distribution by thermal velocity $\b_\m=\b u_\m$ as $\feq_p=\frac{1}{e^{P\cdot\b}+1}$ and $\tfeq_p=\frac{1}{e^{P\cdot\b}-1}$ and the thermal shear is given by
\begin{align}\label{Sij}
S_{ij}=\frac{1}{2}\(\pd_i\b_j+\pd_j\b_i\)-\frac{1}{3}\d_{ij}\pd\cdot\b.
\end{align}
When only thermal shear is present, we can evaluate the LHS as
\begin{align}\label{shear_grad}
&{\hat p}_i\nabla_i\feq_p=-\feq_p(1-\feq_p)\pd_i\b_j\frac{p_ip_j}{E_p}=-\feq_p(1-\feq_p)S_{ij}I^p_{ij}p,\nonumber\\
&{\hat p}_i\nabla_i\tfeq_p=-\tfeq_p(1+\tfeq_p)\pd_i\b_j\frac{p_ip_j}{E_p}=-\tfeq_p(1+\tfeq_p)S_{ij}I^p_{ij}p,
\end{align}
with $I^p_{ij}=\hp_i\hp_j-\frac{1}{3}\d_{ij}$ being a symmetric traceless tensor defined with 3-momentum $p$. We have also replaced $P_iP_j$ by its traceless part by traceless property of $S_{ij}$. Following the method in \cite{Arnold:2000dr}, we parametrize the deviation of distributions by
\begin{align}
\fneq_p=\feq_p(1-\feq_p){\hat f_p},\quad
\tfneq_p=\tfeq_p(1+\tfeq_p){\hat \tf_p},
\end{align}
with the superscripts $(0)$ and $(1)$ counting the order of gradient. To linear order in gradient, the parametrization adopts simple relations for the collision term
\begin{align}\label{linearize}
&f_pf_k(1-f_{p'})(1-f_{k'})-(p,k\leftrightarrow p',k')=\feq_p\feq_k(1-\feq_{p'})(1-\feq_{k'})({\hat f_p}+{\hat f_k}-{\hat f_{p'}}-{\hat f_{k'}}),\nonumber\\
&f_p\tf_k(1+\tf_{p'})(1-f_{k'})-(p,k\leftrightarrow p',k')=\feq_p\tfeq_k(1+\tfeq_{p'})(1-\feq_{k'})({\hat f_p}+{\hat \tf_k}-{\hat \tf_{p'}}-{\hat f_{k'}}),\nonumber\\
&f_pf_k(1+\tf_{p'})(1+f_{k'})-(p,k\leftrightarrow p',k')=\feq_p\feq_k(1+\tfeq_{p'})(1+\tfeq_{k'})({\hat f_p}+{\hat f_k}-{\hat \tf_{p'}}-{\hat \tf_{k'}}).
\end{align}
By rotational symmetry, we expect
\begin{align}\label{para_fneq}
{\hat f}_p=S_{ij}I^p_{ij}\c(p),\quad
{\hat \tf}_p=S_{ij}I^p_{ij}\g(p).
\end{align}
Using \eqref{linearize} and \eqref{para_fneq}, we obtain a linearized Boltzmann equation from \eqref{Boltzmann}:
\begin{align}\label{chi_gamma}
&-f_p(1-f_p)S_{ij}I^p_{ij}p=-\frac{1}{2}\int_{p',k',k}(2\p)\d^4(P+K-P'-K')\frac{1}{16p_0k_0p_0'k_0'}S_{ij}\times\nonumber\\
&\qquad\qquad\qquad\bigg[|{\cal M}|_{\text{Coul},f}^2\(I^p_{ij}\c_p+I^k_{ij}\c_k-I^{p'}_{ij}\c_{p'}-I^{k'}_{ij}\c_{k'}\)f_pf_k(1-f_{p'})(1-f_{k'})\nonumber\\
&\qquad\qquad\qquad+|{\cal M}|_{\text{Comp},f}^2\(I^p_{ij}\c_p+I^k_{ij}\g_k-I^{p'}_{ij}\g_{p'}-I^{k'}_{ij}\c_{k'}\)f_p\tf_k(1+\tf_{p'})(1-f_{k'})\nonumber\\
&\qquad\qquad\qquad\textcolor{black}{+}|{\cal M}|_{\text{anni},f}^2\(I^p_{ij}\c_p+I^k_{ij}\c_k-I^{p'}_{ij}\g_{p'}-I^{k'}_{ij}\g_{k'}\)f_pf_k(1+\tf_{p'})(1+\tf_{k'})\bigg],\nonumber\\
&-\tf_p(1+\tf_p)S_{ij}I^p_{ij}p=-\frac{1}{2}\int_{p',k',k}(2\p)\d^4(P+K-P'-K')\frac{1}{16p_0k_0p_0'k_0'}S_{ij}\times\nonumber\\
&\qquad\qquad\qquad\bigg[|{\cal M}|_{\text{Comp},\g}^2\(I^p_{ij}\g_p+I^k_{ij}\c_k-I^{p'}_{ij}\c_{p'}-I^{k'}_{ij}\g_{k'}\)\tf_pf_k(1-f_{p'})(1+\tf_{k'})\nonumber\\
&\qquad\qquad\qquad\textcolor{black}{+}|{\cal M}|_{\text{anni},\g}^2\(I^p_{ij}\g_p+I^k_{ij}\g_k-I^{p'}_{ij}\c_{p'}-I^{k'}_{ij}\c_{k'}\)\tf_p\tf_k(1-f_{p'})(1-f_{k'})\bigg],
\end{align}
where we have used short-hand notations $\c_p=\c(p)$ and $\g_p=\g(p)$. $S_{ij}$ is arbitrary, thus we can equate its coefficient on two sides. The resulting tensor equations can be converted to scalar ones by contracting with $I^p_{ij}$. The flavor dependence in the amplitude squares can be expressed in terms of elementary amplitude squares as
\begin{align}\label{M2}
&|{\cal M}|_{\text{Coul},f}^2=2N_f|{\cal M}|_{\text{Coul}}^2\nonumber\\
&|{\cal M}|_{\text{Comp},f}^2=|{\cal M}|_{\text{Comp}}^2,\quad |{\cal M}|_{\text{Comp},\g}^2=2N_f|{\cal M}|_{\text{Comp}}^2\nonumber\\
&|{\cal M}|_{\text{anni},f}^2=\frac{1}{2}|{\cal M}|_{\text{anni}}^2,\quad |{\cal M}|_{\text{anni},\g}^2=N_f|{\cal M}|_{\text{anni}}^2,
\end{align}
with
\begin{align}
&|{\cal M}|_{Coul}^2=8e^4\frac{s^2+u^2}{t^2}\nonumber\\
&|{\cal M}|_{Comp}^2=8e^4\frac{s}{-t}\nonumber\\
&|{\cal M}|_{anni}^2=8e^4\(\frac{u}{t}+\frac{t}{u}\)\nonumber.
\end{align}
The factor $2N_f$ in Coulomb case comes from scattering with $N_f$ fermions and anti-fermions. For scattering between identical fermions, the symmetry factor $\frac{1}{2}$ in the final state is compensated by an identical $u$-channel contribution to the LL accuracy. Similarly the factor $2N_f$ in the Compton case comes from scattering of photon with $N_f$ fermions and anti-fermions. The factor $N_f$ in photon pair annihilation corresponds to $N_f$ possible final states and $\frac{1}{2}$ in fermion pair annihilation is a final state symmetry factor.

The phase space integrations are performed in appendix A. The results turn the linearized Boltzmann equations \eqref{chi_gamma} into
\begin{align}\label{LL}
\feq_p(1-\feq_p)\frac{2p}{3}=&e^4\ln e^{-1}\frac{1}{(2\p)^4}\Big[8N_f\frac{\p^3\cosh^{-2}\frac{\b p}{2}\(6\c_p+p((-2+\b p\tanh\frac{\b p}{2})\c_p'-p\c_p'')\)}{72p^2\b^3}\nonumber\\
&\qquad\qquad\qquad+2\frac{\c_p-\g_p}{p}\frac{\p^2}{8\b^2}\frac{4\p}{3}\feq_p(1+\tfeq_p)\Big]\nonumber\\
\tfeq_p(1+\tfeq_p)\frac{2p}{3}=&e^4\ln e^{-1}\frac{1}{(2\p)^4}4N_f\frac{\g_p-\c_p}{p}\frac{\p^2}{8\b^2}\frac{4\p}{3}\tfeq_p(1-\feq_p).
\end{align}
The second equation of \eqref{LL} is algebraic. It is solved by
\begin{align}\label{sol2}
\frac{\g_p-\c_p}{(2\p)^3}=\frac{1}{e^4\ln e^{-1}}\frac{2\b^2}{\p^2N_f}p^2\frac{1+\tfeq_p}{1-\feq_p}.
\end{align}
The first equation is differential and need to be solved numerically. In the limit $\b p\gg1$, the differential terms are subleading, reducing it to an algebraic equation. Combining with \eqref{sol2}, we find the following asymptotic solution
\begin{align}\label{sol1}
\frac{\c(p\to\infty)}{(2\p)^3}=\frac{1}{e^4\ln e^{-1}}\frac{3(1+2N_f)\b^2p^2}{4\p^2N_f^2}.
\end{align}
We have combined $\c_p$ and $\g_p$ with $\frac{1}{(2\p)^3}$ in \eqref{sol2} and \eqref{sol1}. It is convenient as the same factor will appear in phase space integration measure.
\begin{figure}[htbp]
     \begin{center}
          \includegraphics[height=5cm,clip]{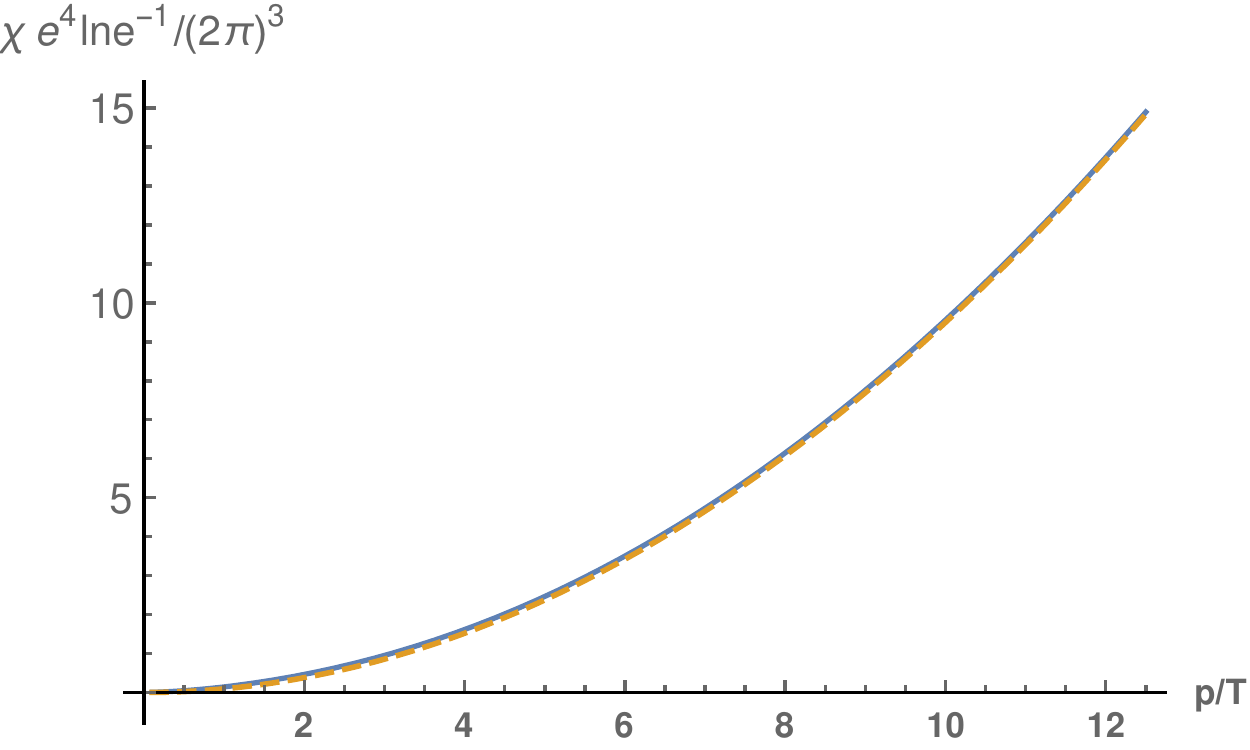}
          \caption{$\c e^4\ln e^{-1}/(2\p)^3$ versus $p/T$ for massless QED with $N_f=2$. Solid and dashed lines correspond to numerical solution and approximate analytic solution \eqref{sol1}. At low $p$, the approximate solution is slightly below the numerical one.}
    \label{fig:chi}
    \end{center}
\end{figure}
The numerical solution is obtained with the boundary condition \eqref{sol1} and $\c(p=0)=0$\footnote{A series analysis of the differential equation in \eqref{LL} around $p=0$ indicate $\c(p)\sim p^2$.}. In fact, it has been pointed out in \cite{Arnold:2000dr} that the ansatz $\c_p,\g_p\sim p^2$ gives very good approximation to the numerical solution.  Fig.~\ref{fig:chi} compares \eqref{sol1} with numerical solution, confirming this point. 
As a further check, we calculate shear viscosity for plasma at constant temperature. In this case $T_{ij}=\h TS_{ij}$. Expressing $T_{ij}$ using kinetic theory, we obtain
\begin{align}
\h=\frac{1}{15}\int_pp\[4N_ff_p(1-f_p)\c_p+2\tf_p(1+\tf_p)\g_p\].
\end{align}
Integrations with numerical solution reproduces the corresponding entries in Table I of \cite{Arnold:2003zc}. Integrations with approximate solution \eqref{sol2} and \eqref{sol1} gives results with an error of about $1\%$ for $N_f=1$ and about $3\%$ for $N_f=2$. We will simply use the approximate solution in the analysis below.

\section{Self-energy correction}

\begin{figure}[htbp]
     \begin{center}
          \includegraphics[height=5cm,clip]{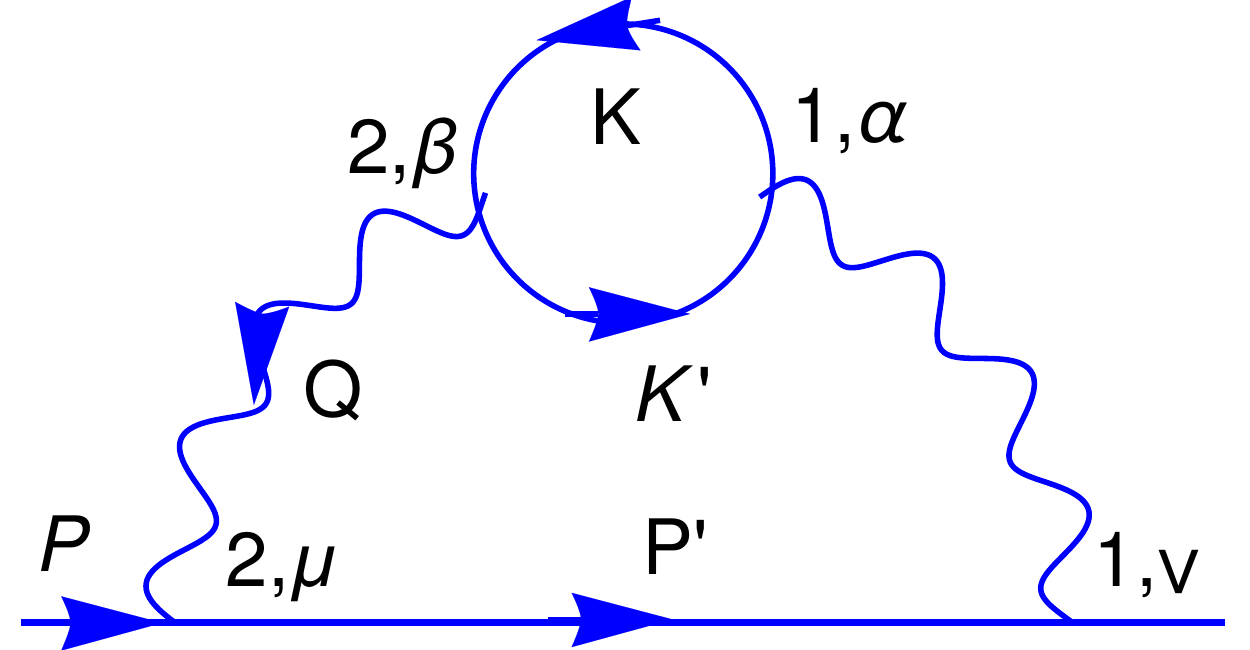}
          \caption{Self-energy of probe fermion from Coulomb scattering with medium fermion. The massive probe fermion carries momentum $P$ and the massless medium fermions run in the loop.}
    \label{fig:Coulomb}
    \end{center}
\end{figure}

In the previous section, we have determined the redistribution of constituents in plasma with thermal shear. Now we introduce a massive probe fermion to the plasma and study its polarization in the shear flow. To this end, we need to calculate self-energy correction to axial component of its Wigner function \eqref{calA}. In general both Coulomb and Compton scatterings contribute to the self-energy\footnote{For probe fermion, pair annihilation is irrelevant.}. Following \cite{Li:2019qkf}, we take the heavy probe limit $m\gg eT$ so that the Coulomb scattering dominates in the self-energy. The Coulomb contribution to the self-energy diagram is depicted in Fig.~\ref{fig:Coulomb}.
We evaluate the self-energy as\footnote{$\S^>(x,y)$ is defined by $-e^2\langle{\slashed A}(x)\ps(x)\bar{\ps}(y){\slashed A}(y)\rangle$.}
\begin{align}\label{Sigma}
\S^>(P)=&{\color{black}+}e^4N_f\int_{P',K',K}(2\p)^4\d^4(P+K-P'-K')\g^\m S^>(P')\g^\n D_{\m\b}^{22}(-Q)D_{\a\n}^{11}(-Q)\nonumber\\
&\times\tr[\g^\a \tS^<(K)\g^\b \tS^>(K')],
\end{align}
with $\int_P=\int\frac{d^4P}{(2\p)^4}$ and $Q=P'-P$. $\S^<$ can be obtained by the replacement $>\leftrightarrow<$, $11\leftrightarrow22$.
The propagators in \eqref{Sigma} are given by
\begin{align}
&S^>(P)=2\p\e(p_0)({\slashed P}+m)(1-f_p)\d(P^2-m^2),\nonumber\\
&\tS^>(K)=2\p\e(k_0){\slashed K}(1-f_k)\d(K^2),\nonumber\\
&D_{\m\b}^{22}(-Q)=\frac{i g_{\m\b}}{Q^2},\quad D_{\a\n}^{11}(-Q)=\frac{-i g_{\a\n}}{Q^2}.
\end{align}
We have indicated propagators of medium fermions by an underline. $\tS^<$ can be obtained by the replacement $1-f_k\to -f_k$. Feynman gauge is used for photon propagators.
The component of self-energy contributing to polarization is $\S^{>\l}=\frac{1}{4}\tr\[\S^>(P)\g^\l\]$. The traces involved in this component are evaluated as
\begin{align}\label{traces}
\tr\[\g^\m S^>(P')\g^\n\g^\l\]=&4\(P'{}^\m g^{\n\l}+P'{}^\n g^{\m\l}-P'{}^\l g^{\m\n}\)2\p\e(p_0')\d(P'{}^2-m^2)(1-f_{p'}),\nonumber\\
\tr\[\g^\a S^<(K)\g^\b S^>(K')\]=&4\(K^\a K'{}^\b+K^\b K'{}^\a-K\cdot K' g^{\a\b}\)(2\p)^2\e(k_0)\e(k_0')\times\nonumber\\
&\d(K^2)\d(K'{}^2)(-f_k)(1-f_k').
\end{align}
Note that the LL contribution arises from the regime $q\ll P,K$, we may replace $\e(p_0')\simeq \e(p_0)=1$ for probe fermion and $\e(k_0)\e(k_0')\simeq \e(k_0)^2=1$. 
Below we assume an equilibrium distribution for probe fermion for illustration purpose. Relaxation of this assumption only involves unnecessary complication. It can be important for realistic modeling of phenomenology, which will be studied elsewhere. The medium fermions is off-equilibrium, with the distribution determined in the previous section.
The combination needed for polarization is $-f_p\S_k^>(P)-(1-f_p)\S_k^<(P)$. Using \eqref{Sigma} and \eqref{traces}, we obtain
\begin{align}\label{fSigma}
&-f_p\S_k^>(P)-(1-f_p)\S_k^<(P)\nonumber\\
=&{\color{black}-}16e^4N_f\int d^3kd^3q\frac{1}{(2\p)^5}\d(p_0+k_0-p_0'-k_0')\frac{1}{8p_0'k_0k_0'}\[2k_kP\cdot K-q_kP\cdot K\]\frac{1}{\(Q^2\)^2}\nonumber\\
&\times\(f_p(1-f_{p'})f_k(1-f_{k'})-f_{p'}(1-f_{p})f_{k'}(1-f_{k})\)\nonumber\\
=&{\color{black}-}16e^4N_f\int d^3kd^4q\frac{1}{(2\p)^5}\d(p_0-p_0'+q_0)\d(k_0-k_0'-q_0)\frac{1}{8p_0'k_0k_0'}\[k_kP\cdot K'+k_k'P\cdot K\]\nonumber\\
&\frac{1}{\(Q^2\)^2}S_{ij}\(I^k_{ij}\c_k-I^{k'}_{ij}\c_{k'}\)\feq_p\feq_k(1-\feq_{p'})(1-\feq_{k'})\nonumber\\
\equiv& S_{ij}R_{ijk}.
\end{align}
We have inserted a factor of $2$ corresponding to fermions and anti-fermion in the loop and kept term up to $O(q^2)$ in the square bracket. In the second equality, we have used the assumption that only the distribution of medium fermions is off-equilibrium.
$R_{ijk}$ involves complicated tensor integrals of ${\vk}$ and ${\vk}'$. They are evaluated by first converting to tensor integrals of ${\vq}$ by rotational symmetry and $\d(k_0-k_0'-q_0)$, which correlates $\vk$ and $\vq$. The resulting tensor integrals of $\vq$ are further performed with rotational symmetry and $\d(p_0-p_0'+q_0)$. Details of the evaluation can be found in appendix B. In the end, we find the following component relevant for spin polarization
\begin{align}
{\cal A}^i=2\p\frac{\e^{ijk}p_jR_{mnk}S_{mn}}{2(p_0+m)}\d(P^2-m^2)\simeq{\color{black}-}\frac{1}{p_0+m}(I_2+I_3)\frac{\e^{iml}p_np_lS_{mn}}{p^5}\d(P^2-m^2)C_f,
\end{align}
with
\begin{align}\label{Is}
I_2=&\frac{\p^2\cosh^{-2}\frac{\b p_0}{2}\((15p^4-87p^2p_0^2+72p_0^4)\ln(\frac{p_0-p}{p_0+p})+\frac{8p^5}{p_0}-126p^3p_0+144pp_0^3\)}{72\b}\nonumber\\
&+\frac{3\cosh^{-2}\frac{\b p_0}{2}\((12p^2p_0-12p_0^3)\ln\frac{p_0-p}{p_0+p}+28p^3-\frac{28p^5}{3p_0^2}-24pp_0^2\)\zeta(3)}{8\b^2},\nonumber\\
I_3=&-\frac{\((p^4-9p^2p_0^2+8p_0^4)\ln\frac{p_0-p}{p_0+p}-\frac{38p_0p^3}{3}+16p_0^3p\)\(\p^2-9\tanh\frac{\b p_0}{2}\zeta(3)\)}{4\b\(1+\cosh(\b p_0)\)}.
\end{align}
and $C_f=\frac{3N_f(1+2N_f)}{4\p^2N_f^2}$.
We reiterate that the self-energy correction scales as $\pd\feq$, with the dependence on coupling cancels as follows: $e^4$ from vertices and $\ln e^{-1}$ from LL enhancement combine to give $\frac{1}{\t_R}\sim e^4\ln e^{-1}$, which is canceled by a counterpart in $\fneq\sim \frac{\pd\feq}{e^4\ln e^{-1}}$.

Before closing this section, we wish to comment on the gauge dependence of \eqref{Is}. We illustrate this with a comparison of Feynman gauge and Coulomb gauge. Let us rewrite \eqref{Sigma} as
\begin{align}\label{fSigma_Pi}
\S^>=e^2\int_Q\g^\m S^>(P')\g^\n D_{\m\b}^{22}(-Q)D_{\a\n}^{11}(-Q)\P^{<\a\b}(Q),
\end{align}
with $\P^{<\a\b}(Q)$ being the off-equilibrium photon self-energy.
In the presence of shear flow, the self-energy can be decomposed into four independent tensor structures as
\begin{align}\label{Pi}
\P^{<\a\b}(Q)=P_T^{\a\b}\P_T^<+P_L^{\a\b}\P_L^<+P_{TT}^{\a\b}\P_{TT}^<+P_{LT}^{\a\b}\P_{LT}^<.
\end{align}
Here $P_{T/L}$ are transverse and longitudinal projectors defined by
\begin{align}\label{projectors}
P^{\a\b}_T=P^{\a\b}-\frac{P^{\a\m}P^{\b\n}Q_\m Q_\n}{q^2},\quad P^{\a\b}_L=P^{\a\b}-P^{\a\b}_T,
\end{align}
with $P^{\a\b}=u^\a u^\b-g^{\a\b}$. $P_{TT}^{\a\b}$ and $P_{LT}^{\a\b}$ are emergent projectors owning to the shear flow, which are constructed as\footnote{The obvious structure constructed by sandwiching $S_{\r\s}$ with two $P_L$ is not independent.}
\begin{align}\label{projectors2}
P_{TT}^{\a\b}=P_T^{\a\r}S_{\r\s}P_T^{\s\b},\quad
P_{LT}^{\a\b}=P_L^{\a\r}S_{\r\s}P_T^{\s\b}+(L\leftrightarrow T).
\end{align}
Note that photon self-energy is gauge invariant but propagator is not. Now we illustrate gauge dependence is generically present by using Feynman and Coulomb gauges.

For LL accuracy, we can simply use bare photon propagators in \eqref{fSigma_Pi}. For spacelike momentum $Q$ relevant for our case, we have a simple relation $D^{11}_{\a\b}=-D^{22}_{\a\b}=-iD^R_{\a\b}$. The retarded propagator in Feynman and Coulomb gauges have the following representations
\begin{align}\label{gauges}
\text{Feynman}:&\;D_{\a\b}^R=P_{\a\b}^T\frac{-1}{Q^2}+\(\frac{Q^2}{q^2}u^\a u^\b-\frac{q_0(u^\a Q^\b+u^\b Q^\a)}{q^2}+\frac{Q^\a Q^\b}{q^2}\)\frac{-1}{Q^2}\nonumber\\
\text{Coulomb}:&\;D_{\a\b}^R=P_{\a\b}^T\frac{-1}{Q^2}+\(\frac{Q^2}{q^2}u^\a u^\b\)\frac{-1}{Q^2}.
\end{align}
Using \eqref{Pi} and \eqref{gauges}, we easily seen contribution to $\S^>$ from $\P_T^<$ and $\P_{TT}^<$ are identical in two gauges. For contribution from $\P_L^<$ and $\P_{LT}^<$, we use Ward identity $\P^{\a\b}Q_\a=0$ and transverse conditions $P_{T/L}^{\a\b}Q_a=0$, $P_{T}^{a\\b}u_\a=0$ to find the following structures, which are present only in Feynman gauge
\begin{align}
\P_L^< P^{\a\b}_L u_\a u_\b Q_\m Q_\n,\quad
\P_L^< P^{\a\b}_L u_\a u_\b u_\m Q_\n+(\m\leftrightarrow\n) ,\quad
\P_{LT}^< P^{\a\b}_L u_\a Q_\m P^T_{\b\n}+(\m\leftrightarrow\n).
\end{align}
We have also confirmed the gauge dependence of self-energy contribution by explicit calculations.

\section{Gauge link contribution}

The gauge dependence we found in the previous section should not be a surprise. The reason is the underlying quantum kinetic theory is derived using a gauge fixed propagators. For Wigner function of the probe fermion, its gauge dependence can be removed by inserting a gauge link. If the gauge field in the link is external, i.e. a classical background, the gauge link simply becomes a complex phase. However, when we consider self-energy of fermions arising from exchanging quantum gauge fields, we need to worry about ordering of quantum field operators from expanding the gauge link and interaction vertex. A systematic treatment of the ordering is still not available at present. We will follow a different approach. Since we have already obtained the axial component of Wigner function without gauge link, we will find correction from expanding the gauge link that contributing at the same order.

\begin{figure}[htbp]
     \begin{center}
          \includegraphics[height=1cm,clip]{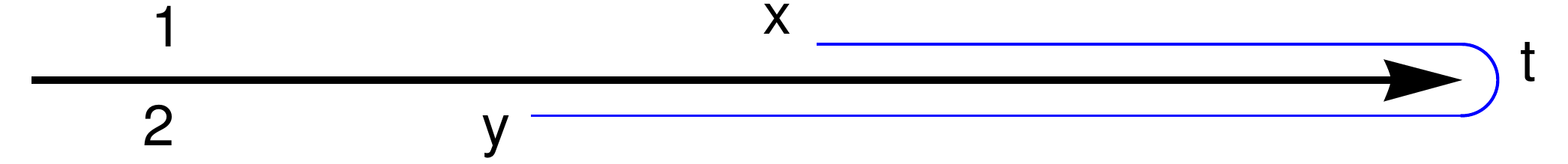}
          \caption{Path for the gauge link in the Schwinger-Keldysh contour. The path in the full spacetime dimension is determined by a straight path joining $x$ and $y$, which is extended to future infinity.}
    \label{fig:path}
    \end{center}
\end{figure}

When fluctuations of quantum gauge fields appear both in the interaction vertices and in the gauge link, it is natural to order them on the Schwinger-Keldysh contour. The latter is also the base of collisional kinetic theory in the recent development of quantum kinetic theory. However we immediately find the well-known straight path for the gauge link becomes inadequate for the Wigner function joining points on forward and backward contours. To find a proper generalization in Schwingwer-Keldysh contour, let us take a close look at the gauge transformation of the bare Wigner function $S^<(x,y)$:
\begin{align}
S^<(x,y)\to e^{-ie\a_2(y)}S^<(x,y)e^{ie\a_1(x)},
\end{align}
with $\a_{1,2}$ being gauge parameters on contour $1$ and $2$ respectively. If there is only classical background field, the gauge fields on contour $1$ and $2$ are the same, we may take $\a_1=\a_2$. In this case, placing the straight path on either contour is equivalent. This is no longer true when quantum fluctuations are present. We propose to use double gauge links
\begin{align}\label{gauge_link}
\bar{S}^<(x,y)=\ps_1(x)\bar{\ps}_2(y)U_2(y,\infty)U_1(\infty,x),
\end{align}
with $U_i(y,x)=\exp\(-ie\int_y^x dw\cdot A_i(w)\)$ and the $i=1,2$ identifying the forward and backward contours respectively. Assuming quantum fluctuations vanishes at past and future infinities, we easily arrive at the gauge invariance of \eqref{gauge_link}. We have not specified the paths for the gauge links appearing in \eqref{gauge_link}. A natural choice would be to take the straight line joining $x$ and $y$ and extending to future infinite. This is illustrated in Fig.~\ref{fig:path}. When there is only classical background gauge field, $A_1=A_2$ so that the two gauge links in \eqref{gauge_link} cancel partially, leaving a phase from the straight path between $x$ and $y$.

\begin{figure}[htbp]
     \begin{center}
          \includegraphics[height=5cm,clip]{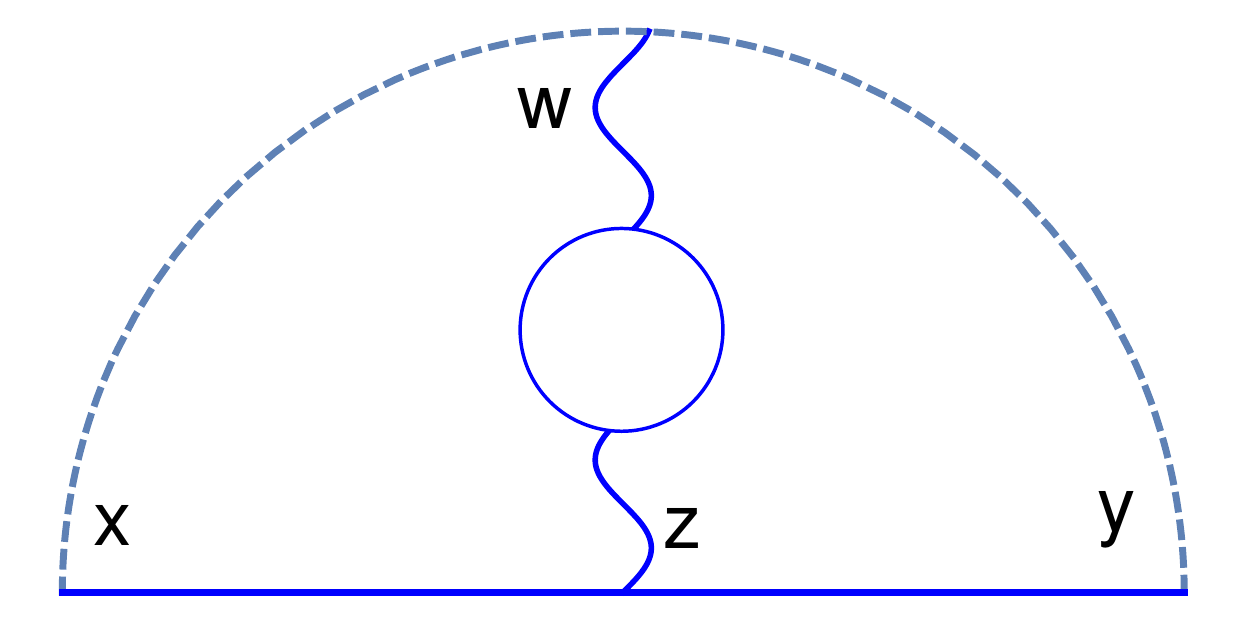}
          \caption{Diagram for gauge link contribution with the propagator connecting one quantum gauge field in the medium and the other in the gauge link. The dashed semi-circle denotes the gauge link. The shear gradient enters through the photon self-energy. In LL approximation, only one insertion of the self-energy is needed.}
    \label{fig:link}
    \end{center}
\end{figure}
Now we are ready to evaluate possible corrections associated with the gauge link. Note that we need a correction of $O(\pd f_0)$. Such a contribution can arise from the diagram in Fig.~\ref{fig:link}. We shall evaluate its contribution to axial component of Wigner function below. Note that the diagram in Fig.\ref{fig:link} contains one quantum fluctuation of gauge field from the link and the other from the interaction vertex. Both fluctuations can occur on either contour $1$ or $2$, and they need to be contour ordered. Enumerating all possible insertions of the two gauge fields along the Schwinger-Keldysh contour, we obtain
\begin{align}\label{link_contour}
&{\color{black}-}e^2S_{11}(x,z)\g^\l S^<(z,y)\(\int_\infty^x dw^\m D_{\l\m}^{11}(z,w)+\int_y^\infty dw^\m D_{\l\m}^<(z,w)\)\nonumber\\
&{\color{black}+}e^2S^<(x,z)\g^\l S_{22}(z,y)\(\int_\infty^x dw^\m D_{\l\m}^>(z,w)+\int_y^\infty dw^\m D_{\l\m}^{22}(z,w)\),
\end{align}
where the two lines corresponding to the vertex coordinate $z$ taking values on contour $1$ and $2$ respectively and the two terms in either bracket corresponding to link coordinate $w$ taking values on contour $1$ and $2$ respectively. The relative sign comes from sign difference of vertices on contour $1$ and $2$. $D_{\l\m}^{>/<}$ stands for resummed photon propagators in medium with shear flow.

Using $S_{11}=-iS_R+S^<$ and $S_{22}=S^<+iS_A$ and the representation
\begin{align}
&S_R=Re S_R+\frac{i}{2}\(S^>-S^<\)\simeq \frac{i}{2}\(S^>-S^<\),\nonumber\\
&S_A=Re S_R-\frac{i}{2}\(S^>-S^<\)\simeq -\frac{i}{2}\(S^>-S^<\),
\end{align}
we obtain $S_{11}\simeq S_{22}\simeq \frac{1}{2}\(S^>+S^<\)$ with $Re S_R$ ignored in the quasi-particle approximation. Similar expressions can be obtained for $D_R$. Plugging the resulting expressions into \eqref{link_contour}, we have
\begin{align}\label{link_expansion}
&{\color{black}-}\frac{e^2}{2}\big[S^>(x,z)\g^\l S^<(z,y)\int_y^x dw^\m D_{\l\m}^<(z,w)-S^<(x,z)\g^\l S^>(z,y)\int_y^x dw^\m D_{\l\m}^>(z,w)\big]\nonumber\\
&{\color{black}-}\frac{e^2}{2}S^<(x,z)\g^\l S^<(z,y)\int_y^x dw^\m\(D_{\l\m}^<(z,w)-D_{\l\m}^>(z,w)\)\nonumber\\
&{\color{black}-}e^2S_{11}(x,z)\g^\l S^<(z,y)\int_\infty^x dw^\m\frac{1}{2}\(D_{\l\m}^>(z,w)-D_{\l\m}^<(z,w)\)\nonumber\\
&{\color{black}-}e^2S^<(x,z)\g^\l S_{22}(z,y)\int_y^\infty dw^\m\frac{1}{2}\(D_{\l\m}^>(z,w)-D_{\l\m}^<(z,w)\).
\end{align}
The first line is very similar to what we have considered in self-energy correction. The other lines are all proportional to the photon spectral density $\r_{\l\m}(z,w)=D_{\l\m}^>(z,w)-D_{\l\m}^<(z,w)$, which is medium independent, thus the other lines are subleading compared to the first one. Below we keep only the first line.

The spin polarization of probe fermion comes from axial component of the Wigner function. We apply Wigner transform to the first line of \eqref{link_expansion}. Since the two terms are simply related by $>\leftrightarrow<$, we focus on the evaluation of the first term. Its Wigner transform is given by
\begin{align}
-\frac{e^2}{2}\int_{s,z,w}e^{iP\cdot s}\int_{P_1,P_2,Q}S^>(P_1)\g^\l S^<(P_2)D_{\l\r}^<(-Q)e^{-iP_1\cdot(x-z)-iP_2\cdot(z-y)+iQ\cdot(z-w)}.
\end{align}
The $z$-integration imposes momentum conservation as $\int_z e^{i(P_1-P_2+Q)\cdot z}=\d(P_1-P_2+Q)$, which allows us to simplify the remaining exponentials as $e^{iP\cdot s}e^{-iP_1\cdot(x-y)+iQ\cdot(y-w)}$. The $w$-integration is performed along the straight line
\begin{align}
\int_y^x dw^\r e^{-iQ\cdot(w-y)}=\int_0^1 dt s^\r e^{-it Q\cdot s}\simeq s^\r,
\end{align}
where we have used $Q\cdot s\ll1$. The condition corresponds to exchange of soft photon, which is necessary for LL enhancement as we already know from the self-energy calculations. We finally replace $s^\r\to-i\frac{\pd}{\pd P_\r}$ to arrive at
\begin{align}
\frac{ie^2}{2}\frac{\pd}{\pd P_\r}\int_QS^>(P)\g^\l S^<(P+Q)D_{\l\r}^<(-Q).
\end{align}
For the axial component, we need the following trace
\begin{align}\label{trace}
\frac{1}{4}\tr\[({\slashed P}+m)\g^\l\({\slashed P}+{\slashed Q}+m\)\g^\m\g^5\]=-i\e^{\a\l\b\m}P_\a Q_\b.
\end{align}
Collecting everything, we obtain the following contributions to axial component of Wigner function
\begin{align}\label{link_exp}
-\frac{e^2}{2}\frac{\pd}{\pd P_\r}&\Big[\int_Q\((1-f_p)f_{p'}D_{\l\r}^>(Q)-f_p(1-f_{p'})D_{\l\r}^<(Q)\)\e^{\a\l\b\m}P_\a Q_\b(2\p)^2\nonumber\\
&\times\d(P^2-m^2)\d(P'{}^2-m^2)\Big],
\end{align}
with $P'=P+Q$. We further use explicit representation of photon propagators in Feynman gauge
\begin{align}\label{D_exp}
D_{\l\r}^>(Q)=(-1)2N_fe^2\int_K\tr[{\slashed K}\g^\a{\slashed K}'\g^\b](1-f_k)(-f_{k'})\frac{-ig_{\l\a}}{Q^2}\frac{ig_{\r\b}}{Q^2}(2\p)^2\d(K^2)\d(K'{}^2),
\end{align}
with $K'=K-Q$.
For the purpose of extracting LL result, we have used bare propagators for photons. The factor of $2N_f$ arises from equal contributions from $N_f$ flavors of fermion and anti-fermion in the medium. A similar expression for $D_{\l\m}^<(Q)$ can be obtained by interchanging $K$ and $K'$ in \eqref{D_exp}. Plugging \eqref{D_exp} into \eqref{link_exp}, we have
\begin{align}\label{link_exp2}
{\color{black}+}N_f\frac{\pd}{\pd P_\r}&\Big[\int\frac{d^3kd^4Q}{(2\p)^52k2p_0'2k'}\(-(1-f_p)f_{p'}(1-f_k)f_{k'}+f_p(1-f_{p'})f_k(1-f_{k'})\)\nonumber\\
&\times4(K_\l K_\r'+K_\r K_\l')\frac{1}{(Q^2)^2}\d(2K\cdot Q)\e^{\a\l\b\m}P_\a Q_\b\d(P^2-m^2)\d(P'{}^2-m^2)\Big].
\end{align}
It has a similar structure with loss and gain terms as the self-energy counterpart \eqref{fSigma}, thus a result proportional to the shear gradient is expected when we take into account redistribution of particles through $f\to \feq+\fneq$.
The remaining task of evaluating the phase space integrals are tedious but straightforward with method sketched in the previous section. Here we simply list the final results with details collected in appendix C
\begin{align}\label{link_final}
{\cal A}^i={\color{black}}\frac{1}{(2\p)}C_f\frac{9\zeta(3)}{2\b^4}(J_1+J_2+J_3+J_4)\frac{\e^{iml}p_np_lS_{mn}}{2p^5}f_p(1-f_{p})\d(P^2-m^2),
\end{align}
with
\begin{align}\label{Js}
&J_1=\frac{8\p\b^2p^3}{p_0},\nonumber\\
&J_2=-\frac{8\p\b^2p^5}{p_0^3},\nonumber\\
&J_3=-\frac{4\p\b^2\(8p^5-56p^3p_0^2+66pp_0^4+(6p^4p_0-39p^2p_0^3+33p_0^5)\ln\frac{p_0-p}{p_0+p}\)}{9p_0^3},\nonumber\\
&J_4=-\frac{-1+e^{\b p_0}}{1+e^{\b p_0}}\frac{2\p\b^3(-2p^2+11p_0^2)\(-4p^3+6pp_0^2+(3p_0^3-3p^2p_0)\ln\frac{p_0-p}{p_0+p}\)}{9p_0^2}.
\end{align}

\section{Discussion}

Let us put together different contributions\footnote{Note that we have assumed the probe fermion has an equilibrium distribution $f_p=\feq_p$.}
\begin{align}\label{As}
&{\cal A}^i_\pd=\frac{2\p}{2(p_0+m)}\e^{iml}p_np_lS_{mn}f_p(1-f_p)\d(P^2-m^2),\nonumber\\
&{\cal A}^i_\S={\color{black}-}C_f\frac{1}{(p_0+m)p^5}\e^{iml}p_np_lS_{mn}(I_2+I_3)\d(P^2-m^2),\nonumber\\
&{\cal A}^i_U={\color{black}}\frac{1}{2\p}\frac{9\z(3)}{2\b^4}C_f\frac{1}{2p^5}\e^{iml}p_np_lS_{mn}(J_1+J_2+J_3+J_4)f_p(1-f_p)\d(P^2-m^2).
\end{align}
The first two lines come from partial derivative and self-energy terms in \eqref{calA} respectively\footnote{In arriving at the first line, an identity similar to \eqref{shear_grad} needs to be used}. The third line comes from the gauge link contribution. The first one is known in the literature \cite{Liu:2021uhn,Becattini:2021suc}. The second and third ones are the main results of the paper. The expressions of $I$ and $J$ can be found in \eqref{Is} and \eqref{Js}.

\begin{figure}[htbp]
     \begin{center}
          \includegraphics[height=5cm,clip]{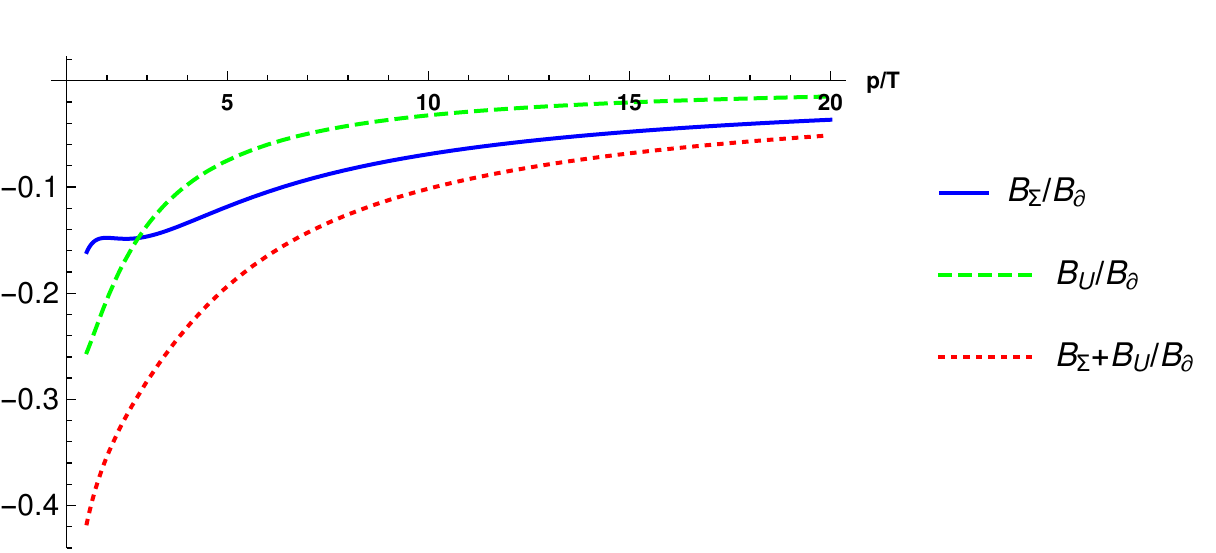}
          \caption{$B_M/B_\pd$ versus $p/T$ for probe fermion mass $m=100$ MeV at $T=150$ MeV for $N_f=2$. $B_M$ are defined by ${\cal A}^i_M=B_M\e^{iml}p_np_lS_{mn}$ with $M=\pd,\S,U$.}
    \label{fig:Bs}
    \end{center}
\end{figure}

It is instructive to take limits to gain some insights from the long expressions. We consider the limit $p_0\gg T$, which allows us to replace in \eqref{Is} the $\cosh$ functions by Boltzmann factors and $\tanh$ function by unity. Similarly $f_p(1-f_p)$ can also be replaced by Boltzmann factor. The limits further allows us to neglect the second line in $I_2$ and $J_1$ through $J_3$. On top of this, we consider separately non-relativistic $m\gg p$ and relativistic limit $m\ll p$. For the former $m\gg p$, we have
\begin{align}\label{massive}
&{\cal A}_\pd^i\simeq\frac{\p}{2m}\e^{iml}p_np_lS_{mn}e^{-\b p_0}\d(P^2-m^2),\nonumber\\
&{\cal A}_\S^i\simeq{\color{black}-}\frac{9\z(3)C_f}{5\b m^2}\e^{iml}p_np_lS_{mn}e^{-\b p_0}\d(P^2-m^2),\nonumber\\
&{\cal A}_U^i\simeq{\color{black}-}\frac{11\z(3)C_f}{5\b m^2}\e^{iml}p_np_lS_{mn}e^{-\b p_0}\d(P^2-m^2).
\end{align}
The fact that the non-relativistic limit is regular in $p$ is a non-trivial: it follows from a cancellation between powers of $p$ from expansion of $I$'s and $J$'s in the numerator and $p^5$ in the denominator in \eqref{massive}, which holds separately for self-energy and gauge link contribution. Since we expect the spin polarization to be well-defined in the non-relativistic limit. The regularity of the results serves as a check of our results.
For the relativistic limit $m\ll p$\footnote{Note that we can still have $m\gg eT$ such that Ignoring Compton scattering is justified.}, we have
\begin{align}\label{massless}
&{\cal A}_\pd^i\simeq\frac{\p}{p}\e^{iml}p_np_lS_{mn}e^{-\b p_0}\d(P^2-m^2),\nonumber\\
&{\cal A}_\S^i\simeq{\color{black}}\frac{(2\p^2-135\z(3))C_f}{9\b p^2}\e^{iml}p_np_lS_{mn}e^{-\b p_0}\d(P^2-m^2),\nonumber\\
&{\cal A}_U^i\simeq{\color{black}-}\frac{9\z(3)C_f}{2\b p^2}\e^{iml}p_np_lS_{mn}e^{-\b p_0}\d(P^2-m^2).
\end{align}
The regularity of the results is also non-trivial in that the logarithmically divergent factor $\ln\frac{p_0-p}{p_0+p}$ as $\frac{p}{m}\to\infty$ is compensated by a vanishing prefactor in both self-energy and gauge link contributions in the relativistic limit.
It is worth mentioning that in both limits ${\cal A}_\S^i$ and ${\cal A}_U^i$ have opposite sign to ${\cal A}_\pd^i$. The magnitude of ${\cal A}_U^i$ is larger(smaller) than ${\cal A}_\S^i$ in the non-relativistic(relativistic) limit. In the limit $p_0\gg T$ we consider, ${\cal A}_\S^i$ and ${\cal A}_U^i$ are suppressed by the factor $\frac{1}{\b m}$ or $\frac{1}{\b p}$ compared to ${\cal A}_\pd^i$. The suppression factor can be easily understood from \eqref{As}: ${\cal A}_\pd^i$ depends on the temperature through the factor $f_p(1-f_p)$, which arises from our local equilibrium assumption on the distribution function of the probe fermion. The other two contributions originate from collisions between probe fermion and medium fermion, thus is characterized by at least one power of temperature, giving rise to a factor $\frac{T}{p_0}$ or $\frac{T}{p}$, which is consistent with the explicit limits we have. The medium dependence is also reflected in the constant $C_f$, which encodes the field content of the medium. In view of application to spin polarization in heavy ion collisions, the contributions from self-energy and gauge link depend on the numerical factors. We plot in Fig.~\ref{fig:Bs} three contributions for phenomenologically motivated parameters, with the caveat that our QED calculation is only meant to provide insights to QCD case. We take $m=100$ MeV, $T=150$ MeV and $p$ in the range of a few GeV. The plot shows for a combined contribution from self-energy and gauge link leads to a modest suppression of the derivative contribution.

\section{Summary and Outlook}

We have revisited spin polarization in a shear flow and found two new contributions. The first one is the self-energy contribution arising from particle redistribution in the shear flow. We illustrate it with a massive probe fermion in a massless QED plasma. It is found that the self-energy contribution is parametrically the same as the derivative contribution considered in the literature.

The self-energy contribution is gauge dependent. In order to restore the gauge invariance of spin polarization, we have proposed a gauge invariant Wigner function, which contains double gauge links stretching along the Schwinger-Keldysh contour. This allows us to include gauge field fluctuations in both forward and backward contours, which is needed for consistent description of gauge field mediated collisions. We have found a second contribution associated with the gauge link, which is also parametrically of the same order.

Both contributions come from particle redistribution in the medium due to the shear flow. The particle redistribution is determined in a steady shear flow, thus the two contributions correspond to non-dynamical ones. A complete description of spin polarization still lacks a dynamical contribution corresponding to the term $a^\m f_A$ in \eqref{calA}. It is worth pointing out that current phenomenological studies seem to indicate an insufficient magnitude from the derivative contribution as compared to measured spin polarization data \cite{Becattini:2021iol,Yi:2021ryh}. The suppression from the new contributions found in this work seems to point to an important role by the dynamical contribution.
Initial efforts have already been made already in \cite{Fang:2022ttm,Wang:2022yli}.

For phenomenological application, several generalizations of the present work are needed: first of all it is crucial to generalize the QED analysis to QCD case. Such a generalization in collisionless limit has been made in \cite{Luo:2021uog,Yang:2021fea}. In the collisional case, we expect the redistribution of both quarks and gluons to play a role; secondly going beyond the LL order is necessary to understand the significance of Compton and annihilation processes in the spin polarization problem; last but not least it is also important to relax our assumption of the equilibrium distribution for the probe fermion. These will be reported elsewhere.

\section*{Acknowledgments}
We are grateful to Jian-hua Gao, Yun Guo and Shi Pu for fruitful discussions. This work is in part supported by NSFC under Grant Nos 12075328, 11735007 (S.L.) and 12005112 (Zy.W.).

\appendix

\section{Phase space integrations in Boltzmann equation}

In this appendix, we perform the phase space integral of the RHS of \eqref{chi_gamma}. As we remarked earlier, the actual integral equation we solve is with $S_{ij}$ replaced by $I_{ij}^p$. We first rewrite the integral measure as
\begin{align}
\int_{p',k',k}(2\p)^4\d^4(P+K-P'-K')=\int\frac{d^3kd^3qdq_0}{(2\p)^6}\d(p_0-p_0'+q_0)\d(k_0-k_0'-q_0),
\end{align}
with $Q=P'-P=K-K'$. We then decompose the vector $\vq$ and $\vk$ as
\begin{align}
&\vq=q\cos\th\hp+\vq_\perp,\quad \vk=k\cos\th'\hp+\vk_\perp,
\end{align}
with $\th$($\th'$) being angles between $\vq$($\vk$) and $\vp$. This allows us to rewrite the integration measure as
\begin{align}
\int d^3kd^3q=\int q^2dqd\cos\th d\ph_{qp}k^2dkd\cos\th'd\ph_{kp},
\end{align}
where $\ph_{qp}$ and $\ph_{kp}$ are azimuthal angles of $\vq$ and $\vk$.

The evaluation of the integral simplifies significantly in the LL approximation, which is known to arise in the region $q\ll p,k$ such that we can perform an expansion in $q$ \cite{Arnold:2000dr}. 
Let us do a power counting in $q$. The two delta functions can be used to eliminate integration of $q_0$ and one factor of $q$, which can be counted effectively as $\frac{1}{q^2}$. The remaining power counting depends on the scattering processes. To be specific, we illustrate with Coulomb scattering amplitude
\begin{align}\label{Coulomb}
|{\cal M}|_{\text{Coul},f}^2=8e^4\frac{s^2+u^2}{t^2}=16e^4\frac{4p^2k^2}{(q_0^2-q^2)^2}\(1-\cos\th'\)^2.
\end{align}
It contains a factor $\frac{1}{q^4}$. On the other hand, the combination $I^p_{ij}\c_p+I^k_{ij}\c_k-I^{p'}_{ij}\c_{p'}-I^{k'}_{ij}\c_{k'}$ can contribute a factor of $q$ as it vanishes in the limit $q_0,q\to 0$. Combining with $q^4$ in the phase space, we obtain an overall power $\frac{1}{q}$. This appears to be more severe than logarithmic divergence. However, we will find an extra factor of $q$ in the actual evaluation. To be safe, we keep correction up to $O(q)$ in the phase space integration.

We will first perform angular integrations using two delta functions, which can be written as
\begin{align}\label{deltas}
&\d(p_0-p_0'+q_0)\simeq\d(-q\cos\th-\frac{q^2}{2p}\sin^2\th+q_0),\nonumber\\
&\d(k_0-k_0'-q_0)\simeq\d(q\cos\O-\frac{q^2}{2k}\sin^2\O-q_0),
\end{align}
where $\O$ is the angle between $\vq$ and $\vk$ and corrections to the arguments higher order in $q$ have been ignored.
We first perform the azimuthal angle integration
\begin{align}\label{delta1}
&\int d\ph_{qp}d\ph_{kp}\d(k_0-k_0'-q_0)=\int d\bar{\ph}d\D\ph\d(q\cos\O-\frac{q^2}{2k}\sin^2\O-q_0)\nonumber\\
\simeq&2\p\frac{2}{q(1+\frac{q_0}{k})}\frac{1}{\(-\cos^2\th'+2\cos\th'\cos\O+1-\cos^2\th-\cos^2\O\)^{1/2}}.
\end{align}
Here $\bar{\ph}$ and $\D\ph$ are the average and difference of $\ph_{qp}$ and $\ph_{kp}$. The delta function fixes $\D\ph$ through $\cos\O=\cos\th\cos\th'+\sin\th\sin\th'\cos\D\ph$.
The square root constrains the integration domain of $\cos\th'$ as: $\cos\th\cos\O-\sin\th\sin\O<\cos\th'<\cos\th\cos\O+\sin\th\sin\O$. The other delta function is easily integrated to give
\begin{align}\label{delta2}
\int d\cos\th\d(p_0-p_0'+q_0)\simeq\frac{1}{q\(1-\frac{q_0}{p}\)}.
\end{align}
Combining \eqref{delta1} and \eqref{delta2} with $\frac{1}{16p_0k_0p_0'k_0'}$, we obtain a simpler expression
\begin{align}
&2\p\frac{2}{q(1+\frac{q_0}{k})}\frac{1}{\(-\cos^2\th'+2\cos\th'\cos\O+1-\cos^2\th-\cos^2\O\)^{1/2}}\frac{1}{q\(1-\frac{q_0}{p}\)}\frac{1}{16p_0k_0p_0'k_0'}\nonumber\\
\simeq&2\p\frac{2}{q^2}\frac{1}{\(-\cos^2\th'+2\cos\th'\cos\O+1-\cos^2\th-\cos^2\O\)^{1/2}}\frac{1}{16p^2k^2}.
\end{align}
It remains to perform the tensor contractions
\begin{align}
&I_{pp}\equiv I_{ij}^p\(I_{ij}^p\c_p-I_{ij}^{p'}\c_{p'}\)=\frac{2}{3}\c_p-\((\hp\cdot\hp')^2-\frac{1}{3}\)\c_{p'},\nonumber\\
&I_{pk}\equiv I_{ij}^p\(I_{ij}^k\c_k-I_{ij}^{k'}\c_{k'}\)=\((\hp\cdot\hk)^2-\frac{1}{3}\)\c_k-\((\hp\cdot\hk')^2-\frac{1}{3}\)\c_{k'}.
\end{align}
Using the following relations
\begin{align}
&\hp\cdot\hp'=\frac{p+q\cos\th}{p+q_0},\quad \hp\cdot\hk=\cos\th',\quad \hp\cdot\hk'=\frac{k\cos\th'-q\cos\th}{k-q_0},
\end{align}
and expanding $\c_{p'}=\c_p+q_0\c_p'+\frac{1}{2}q_0^2\c_p''$, $\c_{k'}=\c_k-q_0\c_k'+\frac{1}{2}q_0^2\c_k''$, we find the following types of integrals
\begin{align}\label{costhp}
\int\cos\th'\frac{\cos^n\th'}{\(-\cos^2\th'+2\cos\th'\cos\O+1-\cos^2\th-\cos^2\O\)^{1/2}},
\end{align}
with $n=0,1,2,3,4$. These integrals evaluate to polynomials in $\cos\th\cos\O$ and $\sin\th\sin\O$, whose values are already fixed by delta functions. We can then perform integrations over $q_0$ and $q$ in order. It turns out that all the potentially $\frac{1}{q}$ divergence vanish after integration over $q_0$. This occurs because the integrand is odd in $q_0$, leaving a logarithmic divergence. The divergence can be rendered finite by screening effect through self-energy of soft photon. Fortunately to extract the LL result, we can simply impose cutoffs in the integral $\int_{eT}^T\frac{dq}{q}$ without explicit inclusion of self-energy, which gives the LL enhancement factor $\ln e^{-1}$ \cite{Arnold:2000dr}. Another significant simplification arises because terms depending on $\c_k$ and its derivatives vanish identically. It follows that the remaining $k$-integration can be performed explicitly, turning the integro-differential equation into a differential equations. We have for the contribution to RHS from Coulomb scattering
\begin{align}
\frac{\p^3\cosh^{-2}\frac{\b p}{2}\(6\c_p+p(-2+\b p\tanh\frac{\b p}{2})\c_p'-p\c_p''\)}{72\b^3p^2}.
\end{align}

\section{Evaluation of self-energy contribution}

We reproduce $R_{mnk}$ defined in \eqref{fSigma} below for convenience
\begin{align}\label{Rmnk}
R_{mnk}=&{\color{black}-}16e^4N_f\int dq_0d^3qd^3k\frac{1}{(2\p)^5}\d(p_0-p_0'+q_0)\d(k_0-k_0'-q_0)[k_kP'\cdot K'+k_k'P'\cdot K]\nonumber\\
&\times\frac{1}{\(Q^2\)^2}\frac{1}{8p_0'k_0'k_0}\(I_{mn}^k\c_k-I_{mn}^{k'}\c_{k'}\)\feq_p\feq_k(1-\feq_{p'})(1-\feq_{k'}).
\end{align}
Defining
\begin{align}
&T_{kmn}=\(k_0k_k'+k_0'k_k\)\(I_{mn}^k\c_k-I_{mn}^{k'}\c_{k'}\),\nonumber\\
&T_{jlmn}=\(k_jk_l'+k_j'k_l\)\(I_{mn}^k\c_k-I_{mn}^{k'}\c_{k'}\),\nonumber
\end{align}
we can rewrite the tensor structures in \eqref{Rmnk} as
\begin{align}
[k_kP'\cdot K'+k_k'P'\cdot K]\(I_{mn}^k\c_k-I_{mn}^{k'}\c_{k'}\)=p_0'T_{kmn}-p_l'T_{klmn}.
\end{align}
By rotational symmetry and the fact the $\vk$ and $\vq$ is correlated by $\d(k_0-k_0'-q_0)$, we can convert $\int d^3k\d(k_0-k_0'-q_0)\(T_{kmn}\;\text{and}\;T_{klmn}\)$ to tensors of $\vq$. Note that $T_{mnk}(T_{klmn})$ is traceless and symmetric in $mn$ and $T_{klmn}$ is also symmetric in $kl$, they can be decomposed using the tensor basis constructed out of $\vq$ with the same symmetry properties as
\begin{align}\label{Tkmn}
&\int d^3k\d(k_0-k_0'-q_0)T_{kmn}=A_3I_{mn}^qq_k+B_3\(q_m\d_{nk}+q_n\d_{mk}-\frac{2}{3}\d_{mn}q_k\),\nonumber\\
&\int d^3k\d(k_0-k_0'-q_0)T_{jlmn}=A_4I_{mn}^qq_jq_l+B_4I_{mn}^qq^2\d_{jl}\nonumber\\
&+C_4\(\frac{1}{2}\(q_m\d_{nj}q_l+q_n\d_{mj}q_l+j\leftrightarrow l\)-\frac{2}{3}\d_{mn}q_jq_l\)+D_4q^2\(d_{jm}\d_{ln}+\d_{jn}\d_{lm}-\frac{2}{3}\d_{mn}\d_{jl}\).
\end{align}
The coefficients are scalar functions of $\vq$, which can be evaluated by contracting \eqref{Tkmn} with the tensor basis
\begin{align}\label{contractions}
\begin{pmatrix}
\frac{2}{3}q^2& \frac{4}{3}q^2\\
\frac{4}{3}q^2& \frac{20}{3}q^2
\end{pmatrix}
\begin{pmatrix}
A_3\\
B_3
\end{pmatrix}
=
\begin{pmatrix}
K_{31}\\
K_{32}
\end{pmatrix},\nonumber\\
\begin{pmatrix}
\frac{2}{3}q^4& \frac{2}{3}q^4& \frac{4}{3}q^4& \frac{4}{3}q^4\\
\frac{2}{3}q^4& 2q^4& \frac{4}{3}q^4& 0\\
\frac{4}{3}q^4& \frac{4}{3}q^4& \frac{14}{3}q^4& \frac{20}{3}q^4\\
\frac{4}{3}q^4& 0& \frac{20}{3}q^4& 20q^4
\end{pmatrix}
\begin{pmatrix}
A_4\\
B_4\\
C_4\\
D_4
\end{pmatrix}
=
\begin{pmatrix}
K_{41}\\
K_{42}\\
K_{43}\\
K_{44}
\end{pmatrix},
\end{align}
with
\begin{align}\label{Ks}
&K_{31}=\int d^3k\d(k_0-k_0'-q_0)\(k_0\vk'\cdot\vq+k_0'\vk\cdot\vq\)\[\((\hk\cdot\hq)^2-\frac{1}{3}\)\c_k-(k\to k')\],\nonumber\\
&K_{32}\simeq\int d^3k\d(k_0-k_0'-q_0)\big[\(4k_0'\vk\cdot\vq-\frac{2}{3}k_0\vk'\cdot\vq-\frac{2}{3}k_0'\vk\cdot\vq\)-(k\leftrightarrow k')\big],\nonumber\\
&K_{41}\simeq\int d^3k\d(k_0-k_0'-q_0)2\vk\cdot\vq\vk'\cdot\vq\[\((\hk\cdot\hq)^2-\frac{1}{3}\)\c_k-(k\to k')\],\nonumber\\
&K_{42}\simeq\int d^3k\d(k_0-k_0'-q_0)2k_0k_0'q^2\[\((\hk\cdot\hq)^2-\frac{1}{3}\)\c_k-(k\to k')\],\nonumber\\
&K_{43}\simeq\int d^3k\d(k_0-k_0'-q_0)\[\(\frac{2}{3}\vk\cdot\vq\,\vk'\cdot\vq+2(\vk\cdot\vq)^2\frac{k_0'}{k_0}\)\c_k-(k\leftrightarrow k')\],\nonumber\\
&K_{44}\simeq\int d^3k\d(k_0-k_0'-q_0)\frac{8}{3}k_0k_0'q^2\[\c_k-\c_{k'}\].
\end{align}
We have again dropped terms higher order in $q$. For later use, we perform a counting of the leading order result of $K$'s. Note that $\vk\cdot\vq\simeq kq\cos\O\simeq kq_0$ and the square brackets are of $O(q_0)$, we deduce $K_{31},K_{32}\sim O(q)$, $K_{41},K_{42},K_{43},K_{44}\sim q_0q$. It follows that to leading order $A_{3},B_{3}\sim O(1/q)$, $A_{4},B_{4},C_{4},D_{4}\sim q_0/q^2$. The explicit results can be obtained by using similar tricks used in appendix A. The expressions are lengthy and not shown here.

For the axial component of Wigner function in \eqref{calA}, we need to integrate the structures $\e^{ijk}p_j\(p_0'T_{mnk}-p_l'T_{klmn}\)$ with $\int dq_0d^3q\d(p_0-p_0'+q_0)$. The tensor integrals can be simplified by noting that the results are expected to be pseudotensors and the only pseudotensor symmetric in $mn$ is $\e^{iml}p_np_l+m\leftrightarrow n$. We can then project the tensor integrands as
\begin{align}\label{pseudotensor}
&\e^{ijk}p_jT_{kmn}=\frac{1}{2p^4}\(\e^{iml}p_np_l+m\leftrightarrow n\)\(T_{kjn}p_jp_np_k-T_{kkn}p_np^2\),\nonumber\\
&\e^{ijk}p_hp_jT_{khmn}=\frac{1}{2p^4}\(\e^{iml}p_np_l+m\leftrightarrow n\)\(T_{khjn}p_hp_jp_np_k-T_{khkn}p_hp_np^2\),
\end{align}
with the understanding that the equal sign hold only after integrating over $\vk$ and $\vq$. Summation over repeated indices is implied. Using \eqref{Tkmn}, we can express the tensor contractions on the RHS of \eqref{pseudotensor} as
\begin{align}\label{Tpp}
&T_{kjn}p_kp_np_j=\vp\cdot\vq\[\(\frac{(\vp\cdot\vq)^2}{q^2}-\frac{1}{3}p^2\)A_3+\frac{4}{3}p^2B_3\],\nonumber\\
&T_{jjn}p_n=\vp\cdot\vq\(\frac{2}{3}A_3+\frac{10}{3}B_3\),\nonumber\\
&T_{khjn}p_hp_kp_np_j=\(\frac{(\vp\cdot\vq)^2}{q^2}-\frac{1}{3}p^2\)\((\vp\cdot\vq)^2A_4+p^2q^2B_4\)+\frac{4}{3}p^2(\vp\cdot\vq)^2C_4+\frac{4}{3}p^2q^2D_4,\nonumber\\
&T_{jhjn}p_hp_n=(\vp\cdot\vq)^2\frac{2}{3}(A_4+B_4)+\(\frac{11}{6}(\vp\cdot\vq)^2+\frac{1}{2}p^2q^2\)C_4+\frac{10}{3}p^2q^2D_4.
\end{align}
With the projection, we can simplify the integral as
\begin{align}
&\e^{ijk}p_jR_{mnk}S_{mn}\nonumber\\
&={\color{black}-}16e^4N_f\int dq_0dqk^2dkd\cos\th'4\p\frac{1}{(2\p)^5}\(p_0'T_{kmn}-p_l'T_{klmn}\)\frac{1}{(Q^2)^2}\frac{1}{8pk^2}\nonumber\\
&\times\frac{1}{\(-\cos^2\th'+2\cos\th'\cos\th\cos\O+1-\cos^2\th-\cos^2\O\)^{1/2}}\feq_p\feq_k(1-\feq_{p'})(1-\feq_{k'}),\nonumber\\
&={\color{black}-}\frac{4}{2\p}I\frac{\e^{iml}p_np_lS_{mn}}{2p^5}C_f,
\end{align}
with the second equality defines $I$. We have also factored out the flavor dependent factors from the overall $N_f$ and $\c$ into the constant $C_f=\frac{3N_f(1+2N_f)}{4\p^2N_f^2}$.

Let us see how logarithmic divergence occurs in $I$ by the following power counting. From the leading order power counting for the coefficients made earlier and using $\vp\cdot\vq\simeq p_0q_0$ from $\d(p_0-p_0'+q_0)$, we deduce the LHS of \eqref{Tpp} are of $\sim O(q_0/q)$\footnote{We have regarded $q_0^2\sim q^2$ and keep only explicit odd power of $q_0$ in the estimate.}. This is to be combined with power counting in the remainder of the integral
\begin{align}\label{q_power}
\frac{q_0}{q}\frac{1}{q}q^4\frac{1}{q^4}\sim \frac{q_0}{q^2},
\end{align}
with the second to fourth factors on the LHS of \eqref{q_power} coming from $\d(p_0-p_0'+q_0)$, $dq_0d^3q$ and $\frac{1}{(Q^2)^2}$ respectively. Similar to the analysis in appendix A, the leading order result vanishes upon integration over $q_0$ because of the oddness of integrand in $q_0$. We need to expand to next to leading order (NLO). It is instructive to split $I$ into three parts:
\begin{align}
&I_1: \e^{ijk}p_jp_0'T_{kmn}\to\e^{ijk}p_jq_0T_{kmn},\quad \e^{ijk}p_jp_l'T_{klmn}\to\e^{ijk}p_jq_lT_{klmn}\nonumber\\
&\text{with LO}\; A_3,B_3,A_4,\dots,D_4,\;\text{and}\; f_pf_k(1-f_{p'})(1-f_{k'})\to f_pf_k(1-f_{p})(1-f_{k}),\nonumber\\
&I_2: \e^{ijk}p_jp_0'T_{kmn}\to\e^{ijk}p_jp_0T_{kmn},\quad \e^{ijk}p_jp_l'T_{klmn}\to\e^{ijk}p_jp_lT_{klmn}\nonumber\\
&\text{with NLO}\; A_3,B_3,A_4,\dots,D_4,\;\text{and}\; f_pf_k(1-f_{p'})(1-f_{k'})\to f_pf_k(1-f_{p})(1-f_{k}),\nonumber\\
&I_3: \e^{ijk}p_jp_0'T_{kmn}\to\e^{ijk}p_jp_0T_{kmn},\quad \e^{ijk}p_jp_l'T_{klmn}\to\e^{ijk}p_jp_lT_{klmn}\nonumber\\
&\text{with LO}\; A_3,B_3,A_4,\dots,D_4,\;\text{and}\; f_pf_k(1-f_{p'})(1-f_{k'})\;\text{expanded to}\; O(q_0).
\end{align}
It turns out $I_1$ vanishes identically. The other two $I$'s are obtained by integration with approximate $\c_k$ from \eqref{sol1}
\begin{align}\label{I_Sigma}
I_2&=\frac{\p^2\cosh^{-2}\frac{\b p_0}{2}\((15p^4-87p^2p_0^2+72p_0^4)\ln(\frac{p_0-p}{p_0+p})+\frac{8p^5}{p_0}-126p^3p_0+144pp_0^3\)}{72\b}\nonumber\\
&+\frac{3\cosh^{-2}\frac{\b p_0}{2}\((12p^2p_0-12p_0^3)\ln\frac{p_0-p}{p_0+p}+28p^3-\frac{28p^5}{3p_0^2}-24pp_0^2\)\zeta(3)}{8\b^2},\nonumber\\
I_3&=\frac{\((p^4-9p^2p_0^2+8p_0^4)\ln\frac{p_0-p}{p_0+p}-\frac{38p_0p^3}{3}+16p_0^3p\)\(\p^2-9\tanh\frac{\b p_0}{2}\zeta(3)\)}{4\b\(1+\cosh(\b p_0)\)}.
\end{align}

\section{Evaluation of gauge link contribution}

Let us define
\begin{align}\label{F_def}
&\frac{\pd}{\pd P_\r}\[\(\feq_p(1-\feq_{p'})D^<_{\l\r}(Q)-\feq_{p'}(1-\feq_{p})D^>_{\l\r}(Q)\)\e^{\a\l\b\m}P_\a Q_\b\d(P^2-m^2)\d(P'{}^2-m^2)\]\nonumber\\
&\equiv\frac{\pd}{\pd P_\r}\[F_\r^\m(P,Q)\d(P^2-m^2)\d(P'{}^2-m^2)\].
\end{align}
We consider $p_0>0$. Since $q\ll p$, we have also $p_0'>0$, allowing us to localize the delta functions in \eqref{F_def} to the particle contributions
\begin{align}
&\frac{\pd}{\pd P_\r}\(\frac{\d(p_0-E_p)}{2E_p}\frac{\d(E_p+q_0-E_{p+q})}{2E_{p+q}}F_\r^\m(p_0=E_p)\)\nonumber\\
=&u^\r\frac{\d'(p_0-E_p)}{2E_p}\frac{\d(E_p+q_0-E_{p+q})}{2E_{p+q}}F_\r^\m(p_0=E_p)+\nonumber\\
&\frac{\pd E_p}{\pd P_\r}\big[-\frac{\d'(p_0-E_p)}{2E_p}\frac{\d(E_p+q_0-E_{p+q})}{2E_{p+q}}-\frac{\d(p_0-E_p)}{2E_p^2}\frac{\d(E_p+q_0-E_{p+q})}{2E_{p+q}}\nonumber\\
&+\frac{\d(p_0-E_p)}{2E_p}\frac{\d'(E_p+q_0-E_{p+q})}{2E_{p+q}}+\frac{\d(p_0-E_p)}{2E_p}\frac{\d(E_p+q_0-E_{p+q})}{2E_{p+q}}\frac{\pd}{\pd p_0}
\big]F_\r^\m(p_0=E_p)\nonumber\\
&-\frac{\pd E_{p+q}}{\pd P_\r}\big[\frac{\d(p_0-E_p)}{2E_p}\frac{\d'(E_p+q_0-E_{p+q})}{2E_{p+q}^2}+\frac{\d(p_0-E_p)}{2E_p}\frac{\d(E_q+q_0-E_{p+q})}{2E_{p+q}^2}\big]F_\r^\m(p_0=E_p)\nonumber\\
&+\frac{\d(p_0-E_p)}{2E_p}\frac{\d(E_q+q_0-E_{p+q})}{2E_{p+q}}P^{\r\l}\frac{\pd}{\pd P_\l}F_\r^\m(p_0=E_p).
\end{align}
The above should be viewed as a function of $p_0$. We find then the term $\propto\d'(p_0-E_p)$ vanishes identically. The remaining terms can be combined by using $\frac{\pd E_p}{\pd P_\r}=\frac{P^{\r\l}P_\l}{E_p}$, $\frac{\pd E_{p+q}}{\pd P_\r}=\frac{P^{\r\l}(P_\l+Q_\l)}{E_{p+q}}$ as
\begin{align}\label{F_split}
&\(\frac{P^{\r\l}P_\l}{E_p}-\frac{P^{\r\l}(P_\l+Q_\l)}{E_{p+q}}\)\frac{\d(p_0-E_p)}{2E_p}\frac{\d'(E_p+q_0-E_{p+q})}{2E_{q}}F_\r^\m(p_0=E_p)\nonumber\\
&-\(\frac{P^{\r\l}P_\l}{E_p^2}+\frac{P^{\r\l}(P_\l+Q_\l)}{E_{p+q}^2}\)\frac{\d(p_0-E_p)}{2E_p}\frac{\d(E_p+q_0-E_{p+q})}{2E_{p+q}}F_\r^\m(p_0=E_p)\nonumber\\
&+\(\frac{P^{\r\l}P_\l}{E_p}\frac{\pd F_\r^\m(p_0=E_p)}{\pd p_0}+P^{\r\l}\frac{\pd F_\r^\m(p_0=E_p)}{\pd P_\l}\)\frac{\d(p_0-E_p)}{2E_p}\frac{\d(E_p+q_0-E_{p+q})}{2E_{p+q}}.
\end{align}
The first two lines and the last line of \eqref{F_split} come from derivatives on $\d(P^2-m^2)$, $\d(P'{}^2-m^2)$ and that on $F_\r^\m$ in \eqref{F_def} respectively. 
From the definition of $F_\r^\m$, it is clear that the last line is nonvanishing only if $\frac{\pd}{\pd P_\r}$ acts on $-\feq_p(1-\feq_{p'})$. Therefore in the last line, we may keep only the corresponding contribution. \eqref{F_split} can be further simplified by noting that with an extra factor of $Q$ in \eqref{F_def} as compared to the self-energy contribution. It is sufficient to approximate factors by their leading order expansion in $Q$. Using $E_{p+q}\simeq E_p$, $\frac{P^{\r\l}P_\l}{E_p}-\frac{P^{\r\l}(P_\l+Q_\l)}{E_{p+q}}=-\frac{P^{\r\l}Q_\l}{E_p}+\frac{P^{\r\l}P_\l\vp\cdot\vq}{E_p^3}$ and integrating by part, we obtain
\begin{align}\label{F_merge}
\bigg[\(\frac{P^{\r\l}Q_\l}{E_p}-\frac{P^{\r\l}P_\l\vp\cdot\vq}{E_p^3}\)\frac{\pd F_\r^\m}{\pd q_0}-\frac{2P^{\r\l}P_\l}{E_p^2}F_\r^\m+\frac{P^{\r\l}P_\l}{E_p}\frac{\pd F_\r^\m}{\pd p_0}\bigg]\d(P^2-m^2)\d(P'{}^2-m^2),
\end{align}
with the understanding that the derivative $\frac{\pd}{\pd p_0}$ acting on $-\feq_p(1-\feq_{p'})$ inside $F_\r^\m$ only. We have also replaced $\frac{\d(p_0-E_p)}{2E_p}\frac{\d(E_p+q_0-E_{p+q})}{2E_{p+q}}$ by $\d(P^2-m^2)\d(P'{}^2-m^2)$. \eqref{F_merge} can be evaluated by the same method discussed in appendix B. We shall not spell out details here but just stress a subtle point related to $\frac{\pd}{\pd q_0}$: As before, we will replace $\vp\cdot\vq$ by $p_0q_0$. It becomes ambiguous whether the replacement should be made before or after the $q_0$-derivative. The correct way is to first replace in all possible places and then take the derivative. The reason is that the projection onto the pseudotensor in \eqref{pseudotensor} is justified only after angular integrations, which imposes $\vp\cdot\vq=p_0q_0$.

Taking $\m=i$ and factoring out the flavor dependent constant $C_f$ as before, we obtain the following results
\begin{align}\label{F_int}
&\int_Q\frac{\pd}{\pd P_\r}F_\r^i\d(P^2-m^2)\d(P'{}^2-m^2)\nonumber\\
&=\int dk\frac{1}{p}\frac{2}{(2\p)^4}L\frac{\e^{iml}p_np_lS_{mn}}{2p^4}\feq_p(1-\feq_{p'})C_f\d(P^2-m^2),
\end{align}
with $L=L_1+L_2+L_3+L_4$ corresponding to four terms in \eqref{F_merge} respectively. The explicit expressions are given below
\begin{align}\label{Ls}
&L_1=\frac{8\p\b^2k^3p^3}{p_0},\nonumber\\
&L_2=-\frac{8\p\b^2k^3p^5}{p_0^3},\nonumber\\
&L_3=-\frac{4\p\b^2k^3\(8p^5-56p^3p_0^2+66pp_0^4+(6p^4p_0-39p^2p_0^3+33p_0^5)\ln\frac{p_0-p}{p_0+p}\)}{9p_0^3},\nonumber\\
&L_4=-\frac{2\p\b^3\tanh\frac{\b p_0}{2}k^3(-2p^2+11p_0^2)\(-4p^3+6pp_0^2+(3p^2p_0+3p_0^3)\ln\frac{p_0-p}{p_0+p}\)}{9p_0^2}.
\end{align}
The $k$-integrals are easily performed to give \eqref{link_final}.

\bibliographystyle{unsrt}
\bibliography{shear_collision}

\end{document}